\documentclass[preprint,12pt]{elsarticle}

\usepackage{amssymb}
\usepackage{amsmath}
\usepackage{algorithm}
\usepackage{algpseudocode}

\usepackage{braket}

\journal{Nuclear Physics B}

\begin{document}

\begin{frontmatter}



\title{Efficient quantum machine learning with inverse-probability algebraic corrections}


\author{J. Seo} 

\affiliation{organization={Department of Physics, Chung-Ang University},
            city={Seoul 06974},
            country={Republic of Korea}}

\begin{abstract}
Quantum neural networks (QNNs) provide expressive probabilistic models by leveraging quantum superposition and entanglement, yet their practical training remains challenging due to highly oscillatory loss landscapes and noise inherent to near-term quantum devices. Existing training approaches largely rely on gradient-based procedural optimization, which often suffers from slow convergence, sensitivity to hyperparameters, and instability near sharp minima. In this work, we propose an alternative inverse-probability algebraic learning framework for QNNs. Instead of updating parameters through incremental gradient descent, our method treats learning as a local inverse problem in probability space, directly mapping discrepancies between predicted and target Born-rule probabilities to parameter corrections via a pseudo-inverse of the Jacobian. This algebraic update is covariant, does not require learning-rate tuning, and enables rapid movement toward the vicinity of a loss minimum in a single step. We systematically compare the proposed method with gradient descent and Adam optimization in both regression and classification tasks using a teacher-student QNN benchmark. Our results show that algebraic learning converges significantly faster, escapes loss plateaus, and achieves lower final errors. Under finite-shot sampling, the method exhibits near-optimal error scaling, while remaining robust against intrinsic hardware noise such as dephasing. These findings suggest that inverse-probability algebraic learning offers a principled and practical alternative to procedural optimization for QNN training, particularly in resource-constrained near-term quantum devices.
\end{abstract}



\end{frontmatter}



\section{Introduction}\label{sec1}

Recent advances in quantum computing have demonstrated that quantum-mechanical properties such as superposition and entanglement can be harnessed to achieve substantial computational advantages over classical approaches. Quantum algorithms for integer factorization and unstructured search have shown polynomial-to-exponential speedups compared with their classical algebraic and logical counterparts, fundamentally reshaping our understanding of efficient computation \cite{shor_1999_qc, prl_2017_grover_algo, prl_2009_hhl_algo, SCHALKERS2024112816}. These developments have motivated a broader investigation into quantum computational models and methods that extend beyond isolated algorithms toward programmable and trainable systems. Within this context, quantum machine learning (QML) \cite{Biamonte2017_qml, Cerezo2022_qml} and, in particular, quantum neural networks (QNNs) have emerged as promising frameworks that combine variational quantum circuits with data-driven optimization \cite{Benedetti_2019_pqc, Hubregtsen2021_pqc, ieee2023_pqc}. QNNs employ the rotating angles of unitary transformations in quantum circuits as trainable parameters. Compared to classical neural networks, QNNs offer access to exponentially large Hilbert spaces even with a modest number of parameters, enabling highly expressive representations of complex functions \cite{Wen2024_CP_qnn_expressivity, Panadero2024_SR_qnn_expressivity, seo_ieee_2025}. Moreover, quantum circuits naturally support non-classical correlations through entanglement, and their unitary structure enforces norm-preserving transformations, which can provide implicit regularization and improved generalization in certain regimes \cite{Coyle2020_npjqi_sampling_efficiency, IEEE2022_quantum_data_compression}.

Despite these conceptual advantages, the practical training of QNNs remains a major challenge. Unlike classical neural networks, where parameters enter the model through multiplicative compositions of weights, QNN parameters typically appear as angles in unitary operators, inducing exponential and highly oscillatory dependencies in the model output. This structural difference gives rise to characteristic optimization pathologies, most notably the barren plateau phenomenon \cite{McClean2018_NC_original_barren_plateau, seo_ieee_2026}, where gradients vanish exponentially with system size or circuit depth. From an optimization perspective, barren plateaus can be interpreted as extensive flat regions or highly localized wells in the loss landscape, caused by the exponential parameterization of unitary transformations rather than by insufficient model capacity. Nevertheless, most existing approaches to QNN training continue to rely on procedural optimization methods, such as gradient descent and its variants, which were originally developed for classical neural networks. These methods update parameters through small, local steps guided by gradient information. In QNNs, however, gradient information is particularly fragile: it can vanish rapidly, become dominated by sampling noise, or fail to provide meaningful guidance toward relevant minima. As a result, procedural learning often exhibits slow convergence and may fail to identify useful solutions even when expressive quantum models are employed.

In contrast, recent work in classical machine learning has revisited ``algebraic'' learning paradigms \cite{kim2025multiplicativelearning}, where parameters are updated through global corrections derived directly from the discrepancy between predicted and target outputs, rather than through incremental gradient-based steps. Such approaches, which include pseudo-inverse and reflection-based methods, treat learning as a problem of aligning model semantics with desired outputs via explicit algebraic projections. These methods have been shown to achieve rapid convergence, reduced sensitivity to hyperparameter tuning, and improved robustness in ill-conditioned learning problems.

The structural properties of QNNs suggest that algebraic learning rules may be particularly well suited to this setting. Even when output gradients are suppressed or corrupted by noise, the discrepancy between predicted and target probabilities can remain significant. This raises the possibility that direct, algebraic corrections in probability space may overcome optimization barriers that hinder procedural learning. In this work, we introduce an algebraic learning framework for QNNs based on inverse-probability mappings, which enables parameters to be updated in a single step toward the vicinity of a loss minimum. We analyze this approach in both regression and classification tasks, and systematically compare its performance with gradient-based methods under finite-sampling and noisy conditions representative of realistic quantum devices. By reframing QNN training as an algebraic problem over probabilistic program semantics rather than as a purely procedural optimization task, this study highlights a complementary pathway for learning in quantum computational models and provides new insights into their robustness and efficiency in practice.

\section{Backgrounds}\label{sec2}

\subsection{Quantum neural networks as probabilistic computational models}\label{sec2.1}

\begin{figure}[t!]
\centering
\includegraphics[width=\linewidth]{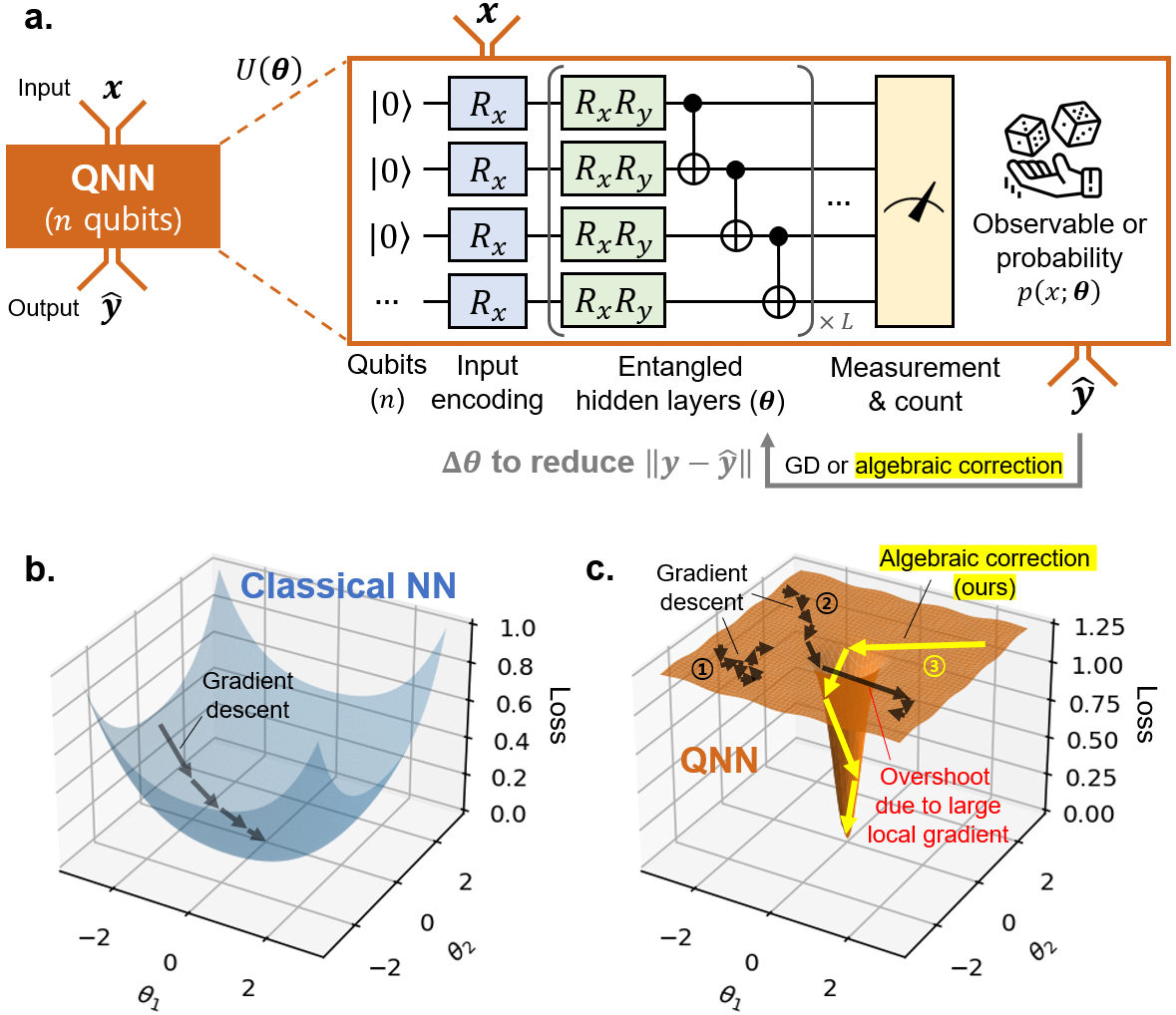}
\caption{Illustration of the quantum neural network and the proposed learning method. \textbf{a}. The architecture of the quantum neural network composed of a variational quantum circuit. \textbf{b}. The description of the gradient descent learning in a classical neural network. \textbf{c}. The description of the gradient descent and the proposed learning methods in a quantum neural network.}\label{fig1}
\end{figure}

Variational quantum circuits (VQCs) are quantum circuits composed of unitary rotations with tunable angle parameters \cite{Benedetti_2019_pqc, Hubregtsen2021_pqc, ieee2023_pqc}. These circuits are the building blocks of QNNs, which enable hybrid quantum-classical neural computing. A QNN, shown in Figure \ref{fig1}a, can be viewed as a parametric probabilistic computation defined by three stages: data embedding $\ket{\psi_0(x)}$, unitary transformation $U(\boldsymbol{\theta})$, and measurement $p(x;\boldsymbol{\theta})$. Here, $x \in \mathbb{R}^K$ is a classical input with $K$ features, which is first embedded into a quantum state via a data-dependent unitary,

\begin{equation}
    \ket{\psi_0(x)} = U_{\text{enc}}(x) \ket{0},
\label{eq_data_embedding} \\
\end{equation}

\noindent where $U_{\text{enc}}(x)$ is typically constructed from single-qubit rotations and feature re-uploading schemes ($R_x$ and $R_y$ in this work). The expressive power of the model is introduced through a trainable unitary operator

\begin{equation}
    U(\boldsymbol{\theta}) = \prod^L_{l=1} U_l(\boldsymbol{\theta}_l),
\label{eq_unitary_transform} \\
\end{equation}

\noindent where each layer $U_l$ consists of parametrized single-qubit rotations and entangling gates ($R_y$, $R_x$, and $CNOT$ gate in this work), and $\boldsymbol{\theta} = \{\boldsymbol{\theta}_l \} \in \mathbb{R}^P$ denotes the full set of trainable parameters.

After applying the unitary evolution, the output of the QNN is obtained via quantum measurement. For a given observable $O$, the model output is defined as an expectation value

\begin{equation}
    p(x;\boldsymbol{\theta}) = \bra{\psi(x;\boldsymbol{\theta})} O \ket{\psi(x;\boldsymbol{\theta})},
\label{eq_measurement} \\
\end{equation}

\noindent where

\begin{equation}
    \ket{\psi(x;\boldsymbol{\theta})} = U(\boldsymbol{\theta}) U_{\text{enc}}(x) \ket{0}.
\label{eq_quantum_state} \\
\end{equation}

In this work, $O$ is chosen as a projector onto a computational basis state, $\ket{1 \dots 1}$, and $p(x;\boldsymbol{\theta})$ corresponds directly to a Born-rule probability and lies in the interval $[0, 1]$. From this perspective, a QNN defines a probabilistic program whose semantics are determined by unitary transformations and whose outputs are expectations over quantum observables.

\subsection{Procedural learning via gradient descent and its limitations}\label{sec2.2}

In most existing approaches, QNN parameters $\boldsymbol{\theta}$ are trained using classical procedural learning rules, most notably gradient descent and its variants. Given a dataset $\{ (x_i, y_i) \}_{i=1}^N$ and a loss function $\mathcal{L}(p, y)$, the parameter update rule is

\begin{equation}
\begin{split}
    \boldsymbol{\theta}_{t+1} &= \boldsymbol{\theta}_t - \eta \nabla_{\boldsymbol{\theta}} \bar{\mathcal{L}} \\
    &= \boldsymbol{\theta}_t - \eta \nabla_{\boldsymbol{\theta}} \left[ \frac{1}{N} \sum^N_{i=1} \mathcal{L}\left(p(x_i;\boldsymbol{\theta}), y_i\right) \right],
\label{eq_gradient_descent} \\
\end{split}
\end{equation}

\noindent where $\eta>0$ is the learning rate ($\eta = 0.1$ in the experiments in Section \ref{sec3}) and $\bar{\mathcal{L}}$ is the averaged loss value for the given dataset. Using the chain rule, the gradient can be written as 

\begin{equation}
    \nabla_{\boldsymbol{\theta}} \bar{\mathcal{L}} = \left( \frac{\partial p}{\partial \boldsymbol{\theta}} \right)^\top \frac{\partial \bar{\mathcal{L}}}{\partial p},
\label{eq_gradient_loss} \\
\end{equation}

\noindent where the Jacobian $J=\partial p/\partial \boldsymbol{\theta}$ encodes the sensitivity of the model output with respect to the parameters. In a QNN, $J$ can be obtained by parameter-shift rules \cite{Mitarai_PRA2018_param_shift, Schuld_PRA2019_param_shift}, which enable the model training (Equation \ref{eq_gradient_descent}) from the given dataset.

Although this gradient descent method could have effectively optimized the parameters of classical neural networks as shown in Figure \ref{fig1}b, in QNN optimization, however, a key difficulty arises from the structure of $J$. Because parameters enter the quantum model through exponentiated unitary operators, the expectation value $p(x;\boldsymbol{\theta})$ becomes a highly oscillatory function of $\boldsymbol{\theta}$. For sufficiently large circuits with $n$ qubits, it has been shown that

\begin{equation}
    \mathbb{E} \left[ \frac{\partial p}{\partial \theta_j} \right]=0, \ \ \  \text{Var}\left[ \frac{\partial p}{\partial \theta_j} \right] \sim \mathcal{O}(e^{-\alpha n}).
\label{eq_gradient_loss} \\
\end{equation}

This phenomenon, known as the barren plateau described in Figure \ref{fig1}c, implies that gradient information becomes exponentially suppressed in a large QNN, even if a discrepancy still remains between the model output and the target value \cite{McClean2018_NC_original_barren_plateau, seo_ieee_2026}. In the presence of finite sampling (shot noise) or quantum hardware noise, the effective gradient is further corrupted, rendering gradient-based procedural updates both slow and unreliable \cite{Wang_NC2021_noise_barren}. As a result, gradient descent may require an impractically large number of iterations or fail to locate meaningful minima altogether, as described with path \textcircled{1} in Figure 1c. Adaptive update schemes such as Adam \cite{kingma2017adam}, which incorporate not only local gradient information but also its historical accumulation, can improve optimization efficiency to some extent. However, in the highly nonlinear loss landscape of QNNs, gradients can increase sharply in the vicinity of a minimum, causing the optimizer to overshoot the optimum, as illustrated by path \textcircled{2} in Figure 1c. This behavior will be discussed in more detail in Section \ref{sec3.3}.

\subsection{Algebraic learning via inverse-probability correction}\label{sec2.3}

In this work, we propose an alternative way to procedural learning, which treats parameter updates as an algebraic correction problem rather than as incremental descent steps. Instead of following the gradient direction, the goal is to directly reduce the discrepancy between predicted ($\hat{y}_i=p(x_i;\boldsymbol{\theta})$) and target ($y_i$) outputs.

First we consider the residual vector $\boldsymbol{r}=\boldsymbol{y}-\boldsymbol{\hat{y}}$. For small parameter updates $\Delta\boldsymbol{\theta}$, a first-order expansion yields

\begin{equation}
    \hat{\boldsymbol{y}}(\boldsymbol{\theta}+\Delta\boldsymbol{\theta}) \approx \hat{\boldsymbol{y}}(\boldsymbol{\theta}) + J\Delta\boldsymbol{\theta}.
\label{eq_first_order} \\
\end{equation}

Our algebraic learning seeks $\Delta\boldsymbol{\theta}$ such that the updated residual ($\boldsymbol{y}-\hat{\boldsymbol{y}}(\boldsymbol{\theta}+\Delta\boldsymbol{\theta}) \approx \boldsymbol{r} - J\Delta\boldsymbol{\theta}$) is minimized in a least-squares sense, with regularization of small parameter updates (Tikhonov regularization \cite{SIAM1999_tikhonov}),

\begin{equation}
    \Delta\boldsymbol{\theta} = \underset{\Delta\boldsymbol{\theta}}{\mathrm{argmin}} \left[ \left\lVert \boldsymbol{r} - J\Delta\boldsymbol{\theta} \right\rVert_2^2 + \lambda \left\lVert \Delta\boldsymbol{\theta} \right\rVert_2^2 \right].
\label{eq_argmin} \\
\end{equation}

The solution is given by the pseudo-inverse form

\begin{equation}
    \Delta\boldsymbol{\theta} = (J^\top J + \lambda I)^{-1} J^\top \boldsymbol{r},
\label{eq_pinv} \\
\end{equation}

\noindent where $\lambda>0$ is a Tikhonov regularization parameter that ensures both numerical stability and small variation of $\boldsymbol{\theta}$.

Crucially, this update rule does not rely on local gradient magnitude alone. Even when individual components of $\nabla_{\boldsymbol{\theta}} \bar{\mathcal{L}}$ vanish, the residual $\boldsymbol{r}$ may remain large, allowing the algebraic update to produce a meaningful parameter correction. This approach is closely related to expectation-reflection methods developed for classical neural networks \cite{kim2025multiplicativelearning}, where multiplicative or inverse mappings are used to align predicted expectations with targets in a single algebraic step. In the quantum setting, the same principle applies, but the correction is performed on the Born-rule probabilities induced by unitary program semantics.

In this work, the term ``inverse-probability'' refers to the inversion of the probabilistic semantics induced by a QNN, which corrects the trainable parameters. Rather than adjusting parameters through local gradient information, the proposed method maps discrepancies in Born-rule probabilities back to parameter updates via a pseudo-inverse of the Jacobian. In this sense, inverse-probability learning addresses the local inverse problem of probabilistic computation.

For probabilistic outputs $p$, it is often advantageous to perform this correction in logit space,

\begin{equation}
    z=\log{\frac{p}{1-p}}, \quad \frac{\partial z}{\partial \boldsymbol{\theta}} = \frac{1}{p(1-p)} \frac{\partial p}{\partial \boldsymbol{\theta}},
\label{eq_logit} \\
\end{equation}

\noindent which can eliminate the nonlinear bounds of the probability space. This logit representation can be useful for classification tasks with cross-entropy loss functions or unbounded regression tasks.

\section{Inverse-probability learning of quantum neural networks}\label{sec3}

In this section, we describe how a QNN can be trained through the inverse-probability algebraic correction introduced in Section \ref{sec2.3}, and compare the resulting learning dynamics with conventional training based on gradient descent. While standard approaches update parameters using local gradient information, our method computes parameter increments so as to directly reduce the mismatch between predicted and target probabilities in a single global step.

\begin{algorithm}[t]
\caption{Inverse-probability algebraic learning for 2-qubit QNNs}
\label{alg1}
\begin{algorithmic}[1]

\Require
Dataset $\mathcal{D} = \{(x_i, y_i)\}_{i=1}^N$,
circuit depth $L$,
shots $S$,
steps $T$,
regularization $\lambda > 0$,

\Ensure
Trained QNN parameters $\boldsymbol{\theta}_T$

\State Initialize parameters $\boldsymbol{\theta}_0 \sim \mathcal{N}(0,\sigma^2)$
\State Set $P = 4 L$ \Comment{Number of trainable parameters}

\For{$t = 0, \dots, T-1$}

    \For{$i = 1, \dots, N$} 
        \State $p_i = p(x_i;\boldsymbol{\theta}_t)$
        \State $\tilde p_i = {k_i}/{S}, \quad
        k_i \sim \mathrm{Binomial}(S, p_i)$
        \State $\hat y_i \leftarrow \mathrm{clip}(\tilde p_i, \varepsilon, 1-\varepsilon)$ \Comment{Forward pass via measurements}
    \EndFor
    
    \For{$j = 1,\dots,P$}
        \State $\boldsymbol{\theta}^{\pm} = \boldsymbol{\theta}_t \pm \frac{\pi}{2} \mathbf{e}_j $ 
        \State $\tilde p_i^{\pm} = p(x_i;\boldsymbol{\theta}^{\pm}) \quad$ (\text{$S$ shots}) 
        \State $J_{ij} = \frac{1}{2}\left(\tilde p_i^{+} - \tilde p_i^{-}\right)$ \Comment{Jacobian via parameter-shift}
    \EndFor

    \State $\mathbf{r}=[r_1, \dots, r_{N}]^\top, \quad r_i = y_i - \hat y_i$
    \State $A = J^\top J + \lambda I, \quad
    \mathbf{b} = J^\top \mathbf{r}$
    \State $\Delta\boldsymbol{\theta}_t = A^{-1}\mathbf{b}$ \Comment{Algebraic parameter correction}

    \State $\boldsymbol{\theta}_{t+1} = \boldsymbol{\theta}_t + \Delta\boldsymbol{\theta}_t$

\EndFor

\State \Return $\boldsymbol{\theta}_T$

\end{algorithmic}
\end{algorithm}

Algorithm \ref{alg1} describes a pseudo-code of the QNN training with the proposed method, inverse-probability algebraic corrections. A key feature of this formulation is that it does not depend on a particular loss function. The parameter correction is defined purely in terms of a discrepancy in probabilistic semantics ($\hat y_i=\tilde p_i$). Therefore, it applies naturally to classification tasks (e.g. using cross-entropy loss where the model output is interpreted as a Bernoulli probability) as well as to regression tasks (e.g. using regression error loss where the target is a continuous normalized quantity in $[0, 1]$). In either case, training proceeds by repeatedly solving an algebraic correction step that reduces the mismatch in the predicted probabilities.

\subsection{Experimental protocal}\label{sec3.1}

For a transparent performance study, we adopt a teacher-student setting \cite{hinton2015distillingknowledgeneuralnetwork, Gou2021_teacher_student} using a minimal two-qubit QNN. The teacher network is a deeper circuit (depth $L_\text{teacher}=6$) used to generate synthetic probabilistic targets $p(x)$ via Born-rule measurement, while the student is a shallower circuit (depth $L_\text{student}=3$) trained to match these targets. For the model output, we evaluate the probability of the $\ket{11}$ outcome,

\begin{equation}
    p(x;\boldsymbol{\theta}) = \left| \braket{11|\psi(x;\boldsymbol{\theta})} \right|^2.
\label{eq_output} \\
\end{equation}

To reflect realistic quantum measurement constraints, we simulate finite-shot sampling. For a true probability $p$, the measured estimate $\tilde p$ is obtained from $S$ shots as

\begin{equation}
    \tilde p = k/S, \quad k \sim \mathrm{Binomial}(S, p),
\label{eq_shot_noise} \\
\end{equation}

\noindent and is numerically stabilized by clipping 

\begin{equation}
    \tilde p \leftarrow \mathrm{clip}(\tilde p, \varepsilon, 1-\varepsilon) \quad (\varepsilon=10^{-6}).
\label{eq_num_stab} \\
\end{equation}

In the following experiment, we use $S=1000$ shots, which will be discussed more in Section \ref{sec3.3}. Since both procedural and algebraic updates depend on sensitivity estimates, we compute $\partial p/\partial \theta_j$ using the parameter-shift rule \cite{Mitarai_PRA2018_param_shift, Schuld_PRA2019_param_shift}, and likewise evaluate these derivatives under the same finite-shot sampling to capture gradient noise induced by sampling. For stability, we apply Tikhonov regularization with $\lambda=0.2$ during the pseudo-inverse estimation.

Gradient descent–based methods require the introduction of a learning-rate hyperparameter $\eta$, and because the optimization scheme in Equation \ref{eq_gradient_descent} is not covariant, a naïve choice of $\eta$ can lead to inefficient training or overshooting of the optimum. In contrast, we emphasize that the proposed inverse-probability learning rule is covariant and does not require a learning-rate hyperparameter. This is because the update is obtained through a first-order algebraic correction that moves the parameters directly toward the vicinity of a local minimum in a single step by Equation \ref{eq_pinv}, thereby providing both improved training efficiency and practical robustness.

\subsection{Comparison of the learning algorithms}\label{sec3.2}

To evaluate the effectiveness of the proposed inverse-probability learning scheme relative to conventional gradient-based optimization, we compare gradient descent (GD), Adam \cite{kingma2017adam}, and the proposed inverse-probability method on the teacher–student benchmark introduced in Section \ref{sec3.1}. In this setting, a shallow student QNN is trained to reproduce the probabilistic outputs generated by a deeper teacher circuit, allowing a controlled assessment of convergence behavior under identical model mismatch and noise conditions.

\begin{figure}[t!]
\centering
\includegraphics[width=\linewidth]{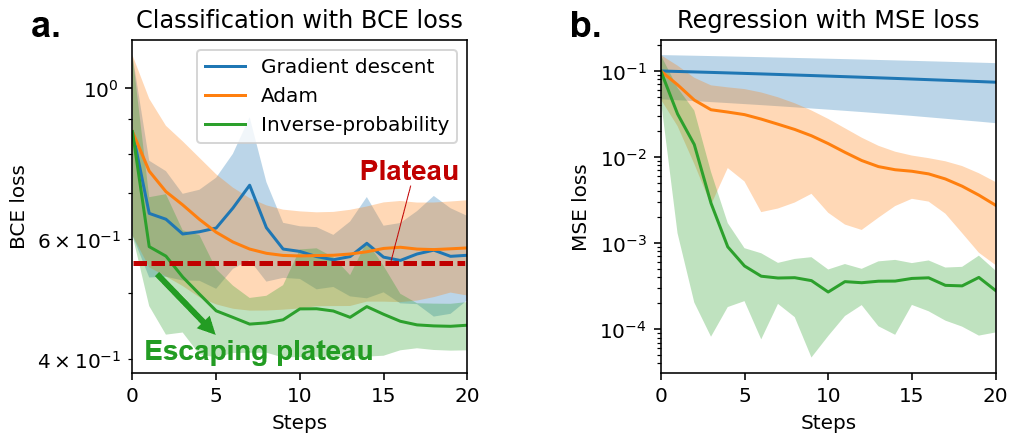}
\caption{Comparison of the gradient descent methods and the proposed learning method. \textbf{a}. Loss history for a classification task with binary cross-entropy (BCE) loss. \textbf{c}. Loss history for a regression task with mean-squared error (MSE) loss. The shaded area indicates the standard deviation of ten ensemble results.}\label{fig2}
\end{figure}

Figure \ref{fig2} presents the training loss history as a function of optimization steps for both classification tasks based on binary cross-entropy (BCE, Figure \ref{fig2}a) loss and regression tasks based on mean-squared error (MSE, Figure \ref{fig2}b) loss. To assess statistical significance and robustness, results are averaged over an ensemble of ten independent random seeds, with the shaded regions indicating the corresponding variability.

Across both classification and regression tasks, GD and Adam exhibit markedly inefficient convergence, often requiring many iterations to achieve modest loss reduction. In contrast, the proposed inverse-probability learning converges rapidly, reaching a low-loss regime within only a few updates. This behavior stems from the algebraic nature of the parameter correction, which directly projects the probability mismatch toward the vicinity of a local minimum in a single step, independently of any learning-rate tuning. As a result, inverse-probability learning achieves both faster convergence and improved robustness compared to procedural gradient-based optimization.

GD and Adam generate small parameter updates at each iteration by exploiting only local gradient information, which typically requires many optimization steps and makes it difficult to locate global minima in complex loss landscapes, as shown in Figure \ref{fig2}. In contrast, our approach constructs the full Jacobian column space from the residual vector over the entire dataset and provides a global algebraic correction of the parameters, via Equation \ref{eq_pinv}. This enables updates toward the vicinity of a minimum in a single step without the need for a learning-rate hyperparameter, resulting in a much faster reduction of the loss compared to conventional methods and escape from loss plateaus shown in Figure \ref{fig2}a. This advantage is particularly relevant for near-term quantum devices, where the number of circuit executions is severely constrained due to the high cost of qubit preparation and coherence maintenance. Under such resource limitations, the proposed method offers enhanced synergy and practical efficiency.

\subsection{Effect of the shot noise in quantum estimation}\label{sec3.3}

Unlike classical neural networks, QNNs possess an intrinsically probabilistic nature that arises naturally from the principles of quantum mechanics, without the need for artificially introduced stochastic mechanisms such as dropout \cite{10.5555/2627435.2670313_dropout} or Bayesian regularization \cite{Sariev01022020_bayesian_regularization}. According to quantum theory, physical observables are not directly accessible as deterministic quantities. Instead, only expectation values of measurement outcomes can be estimated. As a result, the output of a QNN for a given input is obtained by repeatedly measuring a quantum observable and estimating its expectation value through a Monte Carlo sampling process over multiple circuit executions, commonly referred to as ``shots.''

In near-term quantum devices, however, the number of sampling shots $S$ available per inference is severely constrained due to the high cost associated with qubit initialization, control, and coherence preservation \cite{Huang2024}. Consequently, expectation values estimated from a limited number of measurements inevitably deviate from their true values ($\propto 1/\sqrt S$). This statistical fluctuation, known as ``shot noise,'' introduces an additional source of uncertainty into QNN outputs \cite{McClean_2016_shot_noise, Cerezo2021_shot_noise}.

Shot noise poses a significant challenge for conventional QNN training based on gradient descent algorithms. In such procedural optimization schemes, parameter updates rely on local gradient estimates, which are directly computed from noisy expectation values. As a result, even small sampling fluctuations can lead to large variations in the estimated gradients, causing unstable updates, slow convergence, or complete stagnation in the presence of flat or highly structured loss landscapes.

\begin{figure}[t]
\centering
\includegraphics[width=\linewidth]{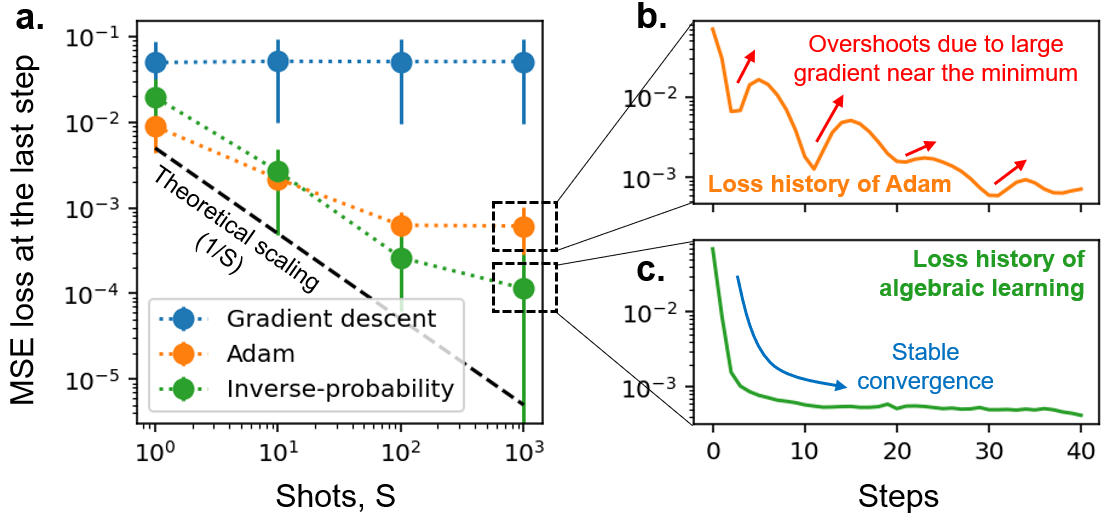}
\caption{Effect of the shot noise in the gradient descent methods and the proposed learning method. \textbf{a}. The final loss values with respect to the number of sampling shots. The vertical errorbars indicate the standard deviations of ten ensemble results. \textbf{b}. A selected loss history for Adam. \textbf{c}. A selected loss history for the proposed learning method.}\label{fig3}
\end{figure}

Figure \ref{fig3}a compares the converged loss values across different optimization algorithms as a function of the number of shots used per estimation in an MSE-based regression task. The comparison highlights how the performance of each method is fundamentally affected by shot noise and by the structure of the QNN loss landscape. A naïve gradient descent scheme, which relies solely on local gradient information, fails to approach the global minimum in this highly flat and oscillatory landscape and exhibits poor convergence across all shot regimes. In the results of both Adam and the inverse-probability method, the predicted error decreases with the number of shots, as expected.

In the extreme low-shot limit $S=1$, the QNN output no longer provides a continuous regression value but instead collapses to binary outcomes of 0 or 1. In this regime, both local gradients and Jacobian estimates become highly noisy, rendering reliable parameter updates difficult. Consequently, both gradient-descent–based methods and Jacobian-based algebraic learning suffer from large errors. Notably, however, Adam exhibits comparatively lower loss than inverse-probability learning in this regime. Although Adam still relies on local gradients, its adaptive update rule, based on accumulated momentum and variance estimates, effectively averages gradient information across multiple optimization steps. This temporal averaging acts as an implicit increase in the effective shot budget, making Adam more robust in single-shot or few-shot settings.

As the number of shots increases beyond $S\sim10$, a clear separation in performance emerges. Algebraic learning rapidly improves and closely follows the theoretical squared-error scaling of $1/S$ (black dashed line in Figure \ref{fig3}a) expected for optimal least-squares estimators. In contrast, Adam deviates from this theoretical scaling and exhibits a slower reduction in loss, even as shot noise is significantly reduced. 

To further analyze this behavior, we examine the loss history for a case with $S=1000$, for Adam (Figure \ref{fig3}b) and algebraic learning (Figure \ref{fig3}c). Algebraic learning exhibits stable and nearly monotonic convergence toward the correct solution, with only small residual oscillations. In contrast, Adam displays non-monotonic behavior characterized by intermittent jumps in the loss. This instability arises from the highly nonlinear and localized structure of the QNN loss landscape. In the loss landscape of QNN, gradients can increase sharply near the global minimum, causing gradient-based updates to overshoot the minimum despite reduced noise levels, as illustrated with path \textcircled{2} in Figure \ref{fig1}c.

While adaptive gradient methods can mitigate overshooting to some extent, they cannot eliminate it entirely in QNNs. As a result, even at large shot numbers, gradient-descent–based methods (GD and Adam) fail to reach the theoretical error floor dictated by shot noise. Taken together, these results indicate a clear regime dependence. For near-term quantum devices limited to single-shot or few-shot estimation, Adam provides more reliable performance, whereas in regimes where tens of shots or more are available, algebraic learning delivers significantly more stable and accurate loss reduction, making it the preferred optimization strategy.

\subsection{Effect of the intrinsic noise in near-term quantum devices}\label{sec3.4}

Beyond finite-shot sampling noise, near-term quantum devices suffer from several sources of intrinsic noise, including decoherence (energy relaxation and dephasing) \cite{EtxezarretaMartinez2021_decoherence, PhysRevX.13.041022_decoherence, Wang2024canerrormitigation_dephasing}, imperfect gate operations \cite{Dalton2024_gate_error, PhysRevA.110.012404_gate_error}, and readout errors \cite{doi:10.1126/sciadv.abi8009_readout_error, PhysRevA.107.062426_readout_error, Cheng2023_nisq}. Unlike shot noise, these hardware-induced errors accumulate with circuit depth and cannot be mitigated by increasing the number of measurements. As a result, deeper quantum circuits experience progressively degraded coherence and entanglement, exacerbating training inefficiencies such as barren plateaus and severely limiting the effective expressivity of QNNs.

To analyze the impact of intrinsic noise on learning performance, we model hardware imperfections through an effective noise channel acting on the quantum state, as described by 

\begin{equation}
    \rho \leftarrow (1-p_\text{deph})\rho + p_\text{deph} Z\rho Z^\dagger,
\label{eq_deph_noise} \\
\end{equation}

\noindent where $\rho$ is the qubit density matrix at each layer, $p_\text{deph} \in [0, 1]$ is the intrinsic dephasing error rate per layer, and $Z$ is the Pauli-Z operator.

Although multiple types of intrinsic noise exist, this study focuses on dephasing-like errors, as dephasing is the dominant factor limiting QNN training performance \cite{Wang2024canerrormitigation_dephasing}. By directly suppressing off-diagonal elements of the density matrix, dephasing destroys quantum coherence and entanglement, which are essential for expressive QNN representations and for maintaining a well-conditioned loss landscape. In this subsection, we ignore the shot noise to investigate the hardware noise in more detail.

\begin{figure}[t]
\centering
\includegraphics[width=\linewidth]{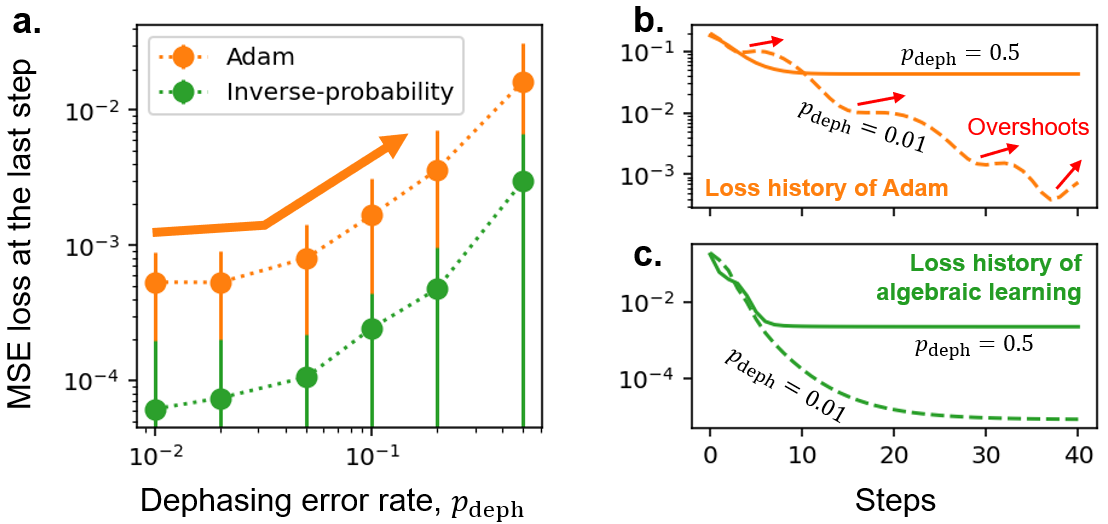}
\caption{Effect of the intrinsic errors in the gradient descent methods and the proposed learning method. \textbf{a}. The final loss values with respect to the dephasing error rates. The vertical errorbars indicate the standard deviations of ten ensemble results. \textbf{b}. Selected loss histories for Adam. \textbf{c}. Selected loss histories for the proposed learning method.}\label{fig4}
\end{figure}

Figure \ref{fig4}a compares the training performance of Adam and the proposed algebraic learning method, with respect to the intrinsic dephasing error rate $p_\text{deph}$. In the low-error regime ($p_\text{deph}<0.05$), algebraic learning exhibits a clear degradation in the converged loss as the error rate increases, reflecting the growing impact of model mismatch induced by intrinsic noise. In contrast, Adam appears to show relatively consistent performance over this range, seemingly insensitive to increases in the error rate. However, this apparent robustness is misleading: for Adam, performance in this regime is already dominated by other limiting factors, such as overshooting near sharp minima discussed in Section \ref{sec3.3}, rather than by intrinsic noise itself. When the dephasing rate exceeds approximately $p_\text{deph}>0.05$, the effect of intrinsic noise becomes dominant even for Adam, and its converged loss increases noticeably with error rate.

This behavior is further clarified in Figure \ref{fig4}b, which shows representative loss histories for Adam under different dephasing levels. At low $p_\text{deph}$, Adam is able to approach the vicinity of the minimum, where overshooting due to the highly nonlinear QNN loss landscape becomes the primary obstacle. In contrast, at higher dephasing rates, intrinsic noise prevents the optimizer from reaching the neighborhood of the minimum altogether, leading to stalled learning and persistently high loss values.

Overall, algebraic learning consistently demonstrates superior training efficiency compared to Adam across all intrinsic error regimes considered, although too high error rate makes both Adam and algebraic learning unable to learn further \cite{Wang_NC2021_noise_barren}. The performance gap between the two methods widens as the intrinsic error decreases, indicating that algebraic learning benefits most from improvements in quantum hardware quality. This trend suggests that as quantum devices continue to reduce coherence loss and gate errors, algebraic learning will increasingly outperform gradient-based optimization for QNN training.

\section{Summary}\label{sec5}

In this study, we reframed the training of quantum neural networks as an algebraic correction problem over probabilistic semantics, rather than a purely procedural optimization task. By exploiting the structure of Born-rule probabilities and their Jacobian with respect to circuit parameters, we derived an inverse-probability learning rule that updates parameters through a global algebraic correction.

Through controlled teacher–student experiments, we demonstrated that conventional gradient-based methods, including Adam, struggle in the highly nonlinear and localized loss landscapes characteristic of QNNs. These methods either stagnate in barren plateaus or overshoot sharp minima, even when shot noise is reduced. In contrast, the proposed algebraic learning method constructs the full Jacobian column space from the residual vector and directly projects parameter updates toward the vicinity of a minimum, leading to rapid and stable convergence.

We further analyzed the effects of finite-shot sampling and intrinsic quantum hardware noise. While single-shot regimes favor adaptive gradient methods due to implicit temporal averaging, algebraic learning quickly surpasses gradient-based approaches as the shot budget increases, following the theoretical optimal error scaling. Under intrinsic dephasing noise, algebraic learning consistently outperforms Adam across all noise levels, with the performance gap widening as hardware quality improves.

Overall, our results highlight inverse-probability algebraic learning as a robust and efficient training paradigm for QNNs. By aligning probabilistic program outputs with target semantics through algebraic inversion, this approach offers a complementary pathway to optimization that is particularly well suited to near-term quantum computing constraints.

\section*{Author contributions}
J.S. contributed to the design of this study, numerical experiments, analyses, and writing the manuscript.

\section*{Data availability}
The codes that generate data in this paper are available at the GitHub repository, https://github.com/jaem-seo/inverse-probability-learning-for-QNN.

\section*{Declaration of competing interest}
The authors declare that they have no known competing financial interests or personal relationships that could have appeared to influence the work reported in this paper.

\section*{Acknowledgments}
This work was supported by a National Research Foundation of Korea (NRF) grant funded by the Korea government (MSIT) (Grant No. RS-2024-00346024).




\bibliographystyle{elsarticle-num} 
\bibliography{mybib.bib}

\end{document}